# Title: Generation of Solar Spicules and Subsequent Atmospheric Heating

**Authors:** Tanmoy Samanta[1], Hui Tian[1]*, Vasyl Yurchyshyn[2], Hardi Peter[3], Wenda Cao[2], Alphonse Sterling[4], Robertus Erdélyi[5,6], Kwangsu Ahn[2], Song Feng[7], Dominik Utz[8], Dipankar Banerjee[9], Yajie Chen[1]

**Affiliations:**

[1] School of Earth and Space Sciences, Peking University, Beijing 100871, People's Republic of China.

[2] Big Bear Solar Observatory, New Jersey Institute of Technology, 40386 North Shore Lane, Big Bear City, CA 92314-9672, USA.

[3] Max Planck Institute for Solar System Research, Justus-von-Liebig-Weg 3, D-37077 Göttingen, Germany.

[4] NASA Marshall Space Flight Center, Huntsville, AL 35812, USA.

[5] Solar Physics and Space Plasma Research Centre, School of Mathematics and Statistics, University of Sheffield, Hounsfield Road, Sheffield, S3 7RH, UK.

[6] Department of Astronomy, Eötvös Loránd University, Budapest, H-1117 Budapest, Hungary.

[7] Faculty of Information Engineering and Automation, Kunming University of Science and Technology, Kunming 650500, People's Republic of China.

[8] Institute for Geophysics, Astrophysics and Meteorology/Institute of Physics, University of Graz, Universitätsplatz 5, 8010 Graz, Austria.

[9] Indian Institute of Astrophysics, Koramangala, Bangalore 560034, India.

*Correspondence to: huitian@pku.edu.cn

**Abstract:** Spicules are rapidly evolving fine-scale jets of magnetized plasma in the solar chromosphere. It remains unclear how these prevalent jets originate from the solar surface and what role they play in heating the solar atmosphere. Using the Goode Solar Telescope at the Big Bear Solar Observatory, we observed spicules emerging within minutes of the appearance of opposite-polarity magnetic flux around dominant-polarity magnetic field concentrations. Data from the Solar Dynamics Observatory showed subsequent heating of the adjacent corona. The dynamic interaction of magnetic fields (likely due to magnetic reconnection) in the partially ionized lower solar atmosphere appears to generate these spicules and heat the upper solar atmosphere.

**One Sentence Summary:** Unprecedented observations with the world's largest Solar Telescope at the Big Bear Solar Observatory reveal that the interaction of entangled solar magnetic fields generates high-speed solar jets, called spicules, and energizes the upper solar atmosphere.

**Main text:** Solar spicules are small-scale, jet-like plasma features observed ubiquitously in the solar chromosphere, the interface between the visible surface (photosphere) of the Sun and its hot outer atmosphere (corona) *(1–4)*. Spicules may play a role in the supply of energy and material to the corona and solar wind *(4, 5)*. They often have lifetimes ranging from 1 to 12 min and are characterized by rising and falling motions with speeds of 15 to 40 km/s *(1, 6)*. Some spicules may have apparent speeds of ~100 km/s and lifetimes less than 1 min *(7)*. In on-disk observations of the chromosphere, these spicules often appear as elongated, short-lived dark structures *(8)*. Some spicules are heated to ≳100,000 K *(9, 10)*.

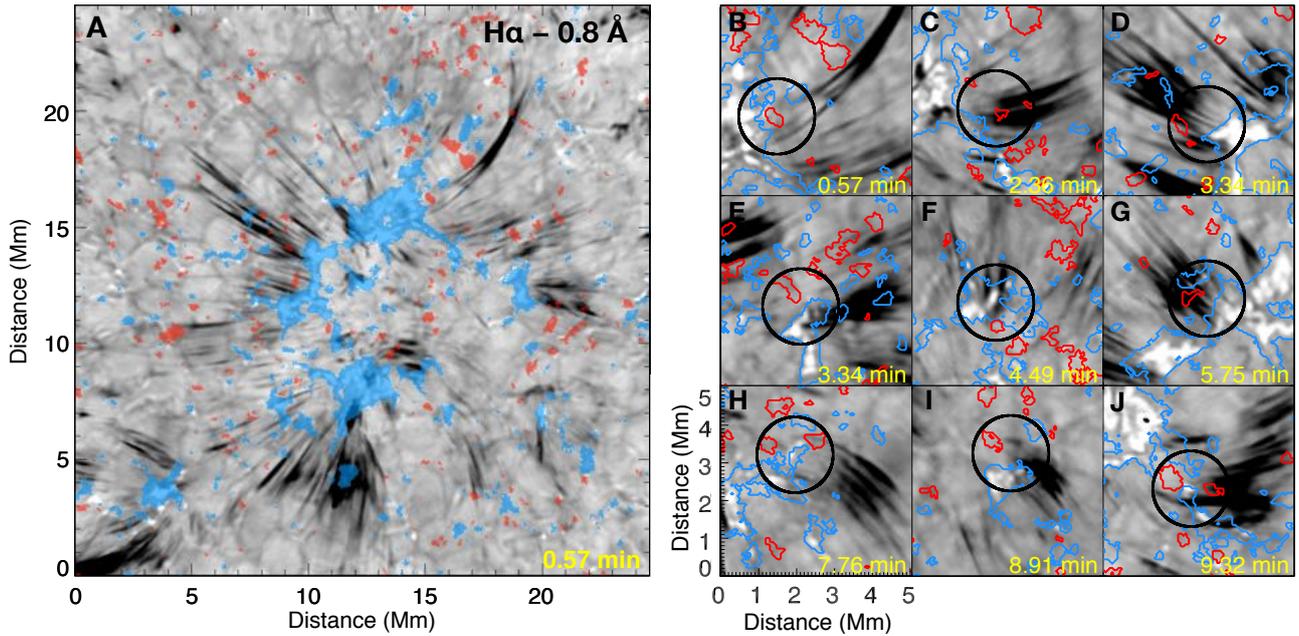

**Fig. 1. Association of enhanced spicular activities with opposite-polarity magnetic fields.** (A) Hα blue wing image (gray-scale) overlain with a binary magnetic field map shown in blue and red, representing longitudinal flux densities of at least +10 Mx/cm² and −10 Mx/cm², respectively (1 Mx = 10⁻⁸ Wb; the unit of Mx/cm² is equivalent to Gauss). Movie S1 shows an animated version of this panel. Figure S1 shows the location of this region, and observational parameters are listed in table S1. (B to J) Examples of enhanced spicular activities. Blue and red contours outline regions of ±10 Mx/cm² for the longitudinal flux density. Axes are the same in different panels. The black circle (with a radius of 1 Mm) in each panel highlights a region around the footpoint of a region of enhanced spicular activity, where at least one small negative-polarity magnetic element is observed in each case.

Theoretical models of spicules have included driving by shock waves *(2, 3)*, Alfvén waves *(11, 12)*, amplified magnetic tension *(13)*, or magnetic reconnection *(14)*. However, observations of their formation process are limited, owing to insufficient resolution and sensitivity. Two observations revealed a tendency for the presence of opposite-polarity magnetic flux near magnetic field concentrations during the occurrence of some spicules *(15, 16)*. However, further analysis did not yield an obvious association between spicules and magnetic field evolution *(16)*.

We observed spicules (fig. S1) using the 1.6-m Goode Solar Telescope (GST) *(17, 18)* at the Big Bear Solar Observatory (BBSO). We performed Hα wing observations and simultaneous magnetic flux measurements with GST's Near Infra- Red Imaging Spectropolarimeter (NIRIS) *(19)*. NIRIS enables us to obtain information on photospheric magnetic fields by spectropolarimetric observations of the Fe I 1.56 mm line *(18)* (figs. S2 and S3). Figure 1A shows a solar image at the blue wing (−0.8 Å from line core) of the Hα line. It is dominated by numerous elongated dark jets, i.e., spicules. These spicules mostly originate from the magnetic network, indicated by the locations of magnetic field concentrations with positive polarity (Figs. 1 and S2).

In addition to frequent individual spicules, occasionally several spicules originate simultaneously from a small region, appearing as enhanced spicular activity at a single location (movie S1). These enhanced spicular activities are accompanied by the presence of weak magnetic elements with a polarity opposite to the dominant polarity of the magnetic network around their footpoints (Fig. 1, B to J). When spicules occur, these weak elements are typically within several hundred kilometers from the edge of the strong network fields. By contrast, the strong and evolving unipolar fields (present for a much longer time in the network) generally do not produce enhanced spicular activities.

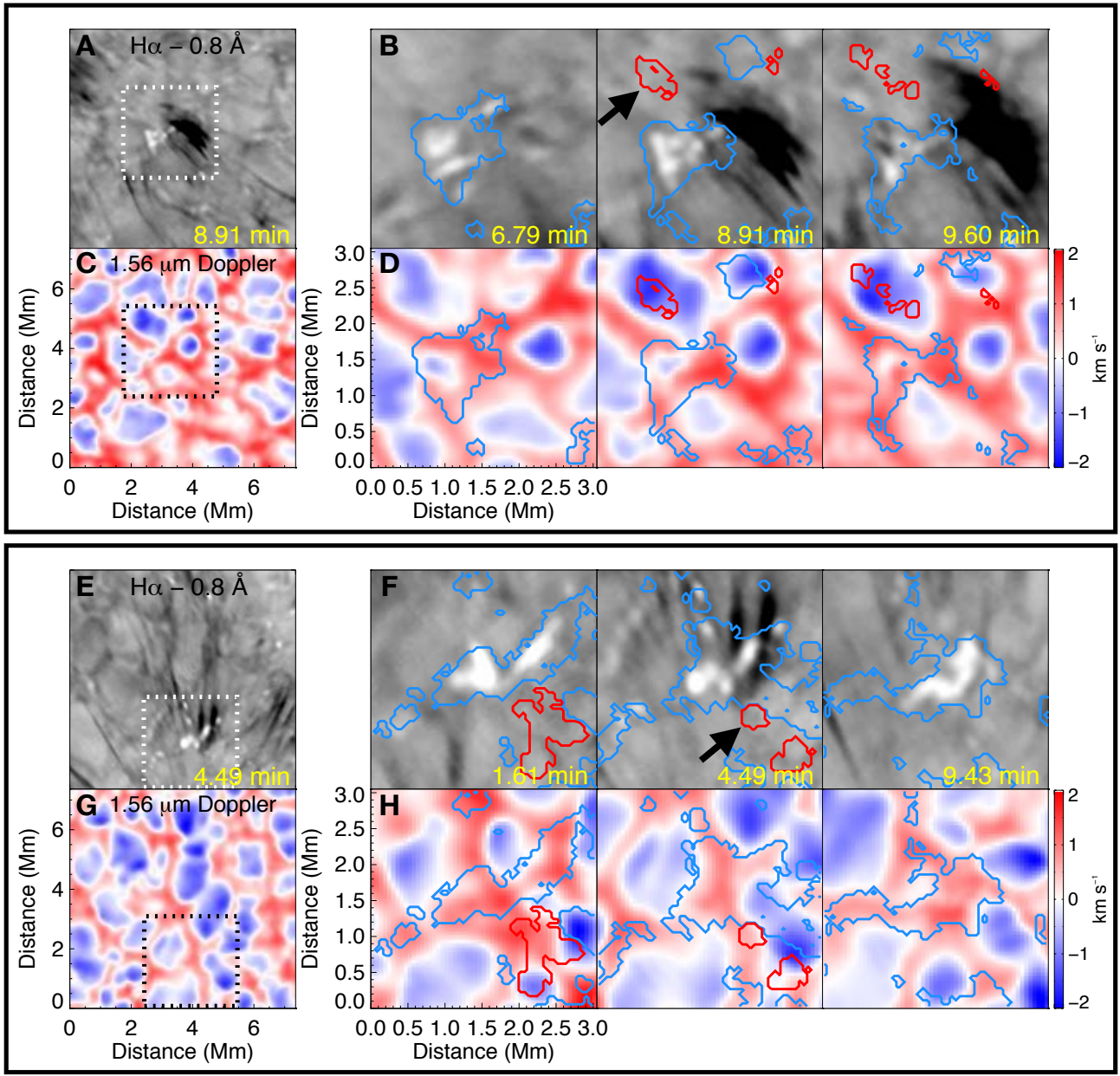

**Fig. 2. Enhanced spicular activity triggered by flux emergence (A to D) (movie S2) and flux cancellation (E to H) (movie S3).** (A) Enhanced spicular activity in a Hα blue wing image. (C) Photospheric Doppler shift pattern of the same region. (B and D) Temporal evolution around the spicule footpoint region [dotted boxes in (A) and (C)]. Contour colors and levels are the same as in Fig. 1B. The arrow in (B) indicates the presence of an opposite-polarity flux. Panels (E) to (H) are the same as (A) to (D) but for a different region.

These enhanced spicular activities appear to be driven by the dynamical interaction of magnetic fields, often preceded by new flux emergence or appearance, and sometimes accompanied by apparent flux cancellation near the network edge. Figure 2, A to D, shows a patch of small-scale weak field with negative polarity that emerges near the strong positive-polarity network fields in the photosphere. Its coincidence with a patch of large blue shift of Fe I also indicates the emergence of the field (movie S2 and figs. S4 and S5). This flux emergence is followed within minutes by enhanced spicular activity, observed in the blue wing of Hα. Figure 2, E to H, shows a larger patch of weak negative-polarity field that approaches the strong network fields; the subsequent flux cancellation leads to enhanced spicular activity (movie S3). The flux cancellation takes place at the boundary of a convection cell that is characterized by red shifts of the Fe I line.

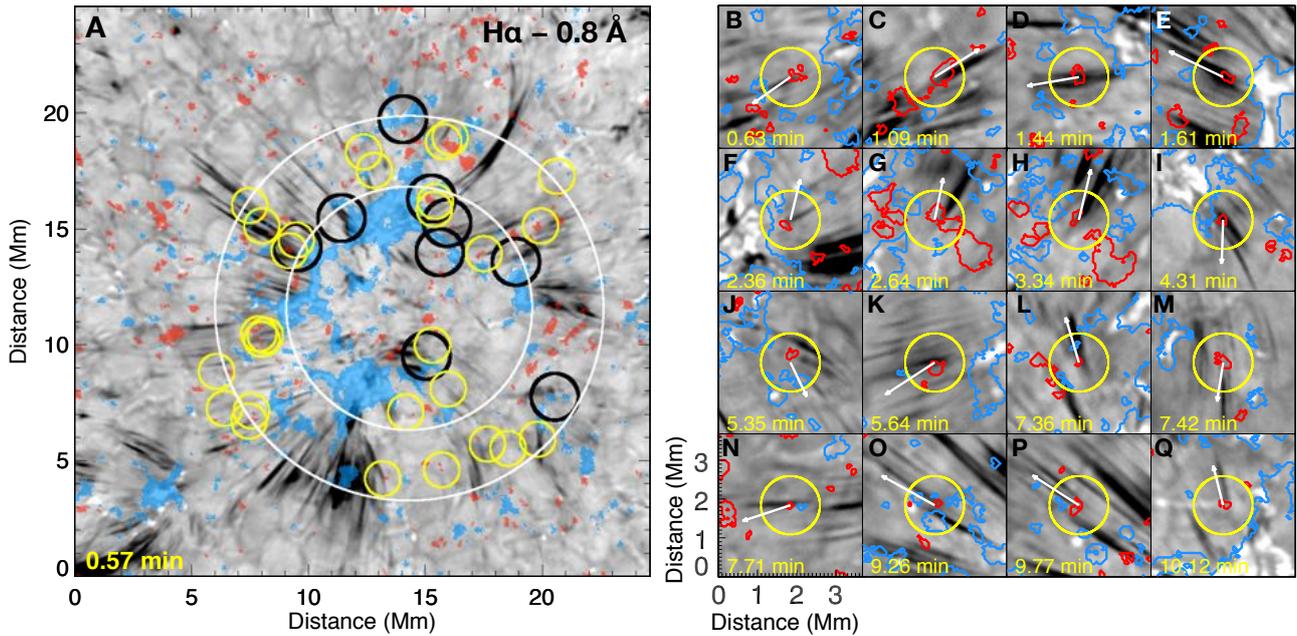

**Fig. 3. Connection of individual spicules to opposite-polarity magnetic fluxes.** (A) The same image as Fig. 1A, overlain with an inner white circle (a radius of 5.25 Mm) representing the approximate boundary of the network, and an outer white circle 3 Mm outside it. Black circles mark the footpoint regions of the same regions with enhanced spicular activity as shown in Fig. 1, B to J. Yellow circles have a radius of 0.75 Mm and indicate regions shown in (B) to (Q) and in fig. S7, which mostly lie within the outer white circle. Movie S4 shows an animated version of this panel. (B to Q) Sixteen examples showing the presence of an opposite-polarity flux near the spicule footpoint (indicated by the yellow circles). Contour colors and levels are the same as in Fig. 1B. The white arrow in each panel indicates the direction radially outward from the center of the white circles.

Almost all the enhanced spicular activities that we observed are associated with emergence or appearance of negative-polarity fluxes and/or subsequent flux cancellation around the boundary of the positive-polarity magnetic network (Figs. 2 and S6 and movie S4). Furthermore, many individual spicules appear to originate, sometimes repeatedly, from small-scale negative-polarity magnetic features located near the strong network fields (Fig. 3 and fig. S7). Although small-scale flux emerges or appears ubiquitously in the quiet Sun, our observations indicate that only when it is close to the strong network fields (often <3 Mm; Fig. 3 and movie S4) does it generate spicules. For some small spicules, no opposite polarity is detected at their footpoints. Because the magnetograms have a spatial resolution (~150 km) about three times lower and a cadence (71 s) around 20 times slower than the Hα images, there may be smaller-scale or highly dynamic fields at the footpoints of these small spicules that we cannot detect.

Our results support the hypothesis that fast spicules originate from magnetic reconnection *(14, 20, 21)*. It is possible that a subphotospheric local dynamo mechanism *(22)* or magneto-convection process *(13, 23)* generates weak magnetic fields close to the large-scale network fields. These small-scale weak fields may occasionally emerge into the photosphere and rise to the chromosphere, where they could reconnect with adjacent or overlying network fields to produce spicules. Alternatively, an opposite-polarity magnetic element could appear as a result of the coalescence and concentration of previously existing dispersed and unresolved fluxes *(24)*, then reconnect with the network fields to generate spicules.

Spicules might supply hot plasma to the solar corona *(4, 5, 9)*. We analyzed coronal observations of the same region with the Atmospheric Imaging Assembly (AIA) *(25)* on the Solar Dynamics Observatory (SDO) spacecraft *(18)*. Most of the enhanced spicular activities are seen to channel hot plasma into the corona (Fig. 4, figs. S8 and S9, and movies S5 to S8). Coronal emission (visible in AIA images at 171 Å) generally appears at the top of the spicules. Our observations in a quiet-Sun region (fig. S1) complement previous observations *(4, 26)* that identified similar coronal signatures for some chromospheric upflow events observed above the solar limb or in on-disk active regions (regions around sunspots). Our observations reveal that magnetic reconnection events at network boundaries can drive spicules and produce hot plasma flows into the corona, providing a link between magnetic activities in the lower atmosphere and coronal heating. It remains unclear whether this process can provide sufficient heating to explain the high temperature of the corona *(27, 28)*.

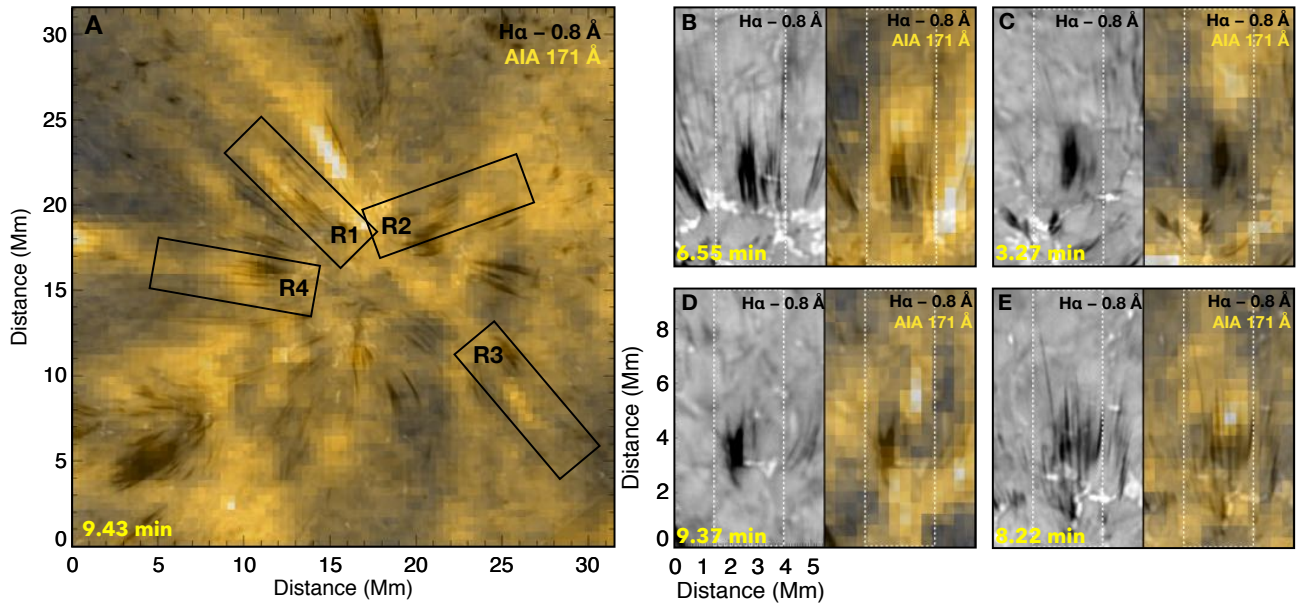

**Fig. 4. Coronal connection of enhanced spicular activities (movie S5).** (A) Hα blue wing image (grayscale) overlain with the simultaneously taken AIA 171 Å image (yellow). Boxes indicate regions shown in the other panels. (B to E) Four examples showing the enhancement of coronal emission above regions of enhanced spicular activity (more examples are in fig. S8). The Hα blue wing and the same image overlain with the simultaneous AIA 171 Å image are shown in each pair of panels. The white dotted boxes in (B) to (E) correspond to the black boxes (R1 to R4) in (A), respectively.

Heated material sometimes falls back from the corona (fig. S10 and movie S9), which could be responsible for the prevalent redshifts of emission lines formed in the chromosphere- corona transition region *(29, 30)*. Our observations of the formation of spicules, the subsequent heating, and the return flows reveal a complete mass cycling process between the chromosphere and corona.

**References and Notes:**

1. J. M. Beckers, *Annu. Rev. Astron. Astrophys* **10**, 73 (1972).
2. A. C. Sterling, *Sol. Phys.* **196**, 79 (2000).
3. B. De Pontieu, R. Erdélyi, S. P. James, *Nature* **430**, 536 (2004).
4. B. De Pontieu, *et al.*, *Science* **331**, 55 (2011).
5. R. G. Athay, T. E. Holzer, *Astrophys. J.* **255**, 743 (1982).
6. G. Tsiropoula, *et al.*, *Space Sci. Rev.* **169**, 181 (2012).
7. B. De Pontieu, et al., *Pub. Astron. Soc. Jpn.* **59**, S655 (2007).
8. L. Rouppe van der Voort, J. Leenaarts, B. De Pontieu, M. Carlsson, G. Vissers, *Astrophys. J.* **705**, 272 (2009).
9. H. Tian, *et al.*, *Science* **346**, 1255711 (2014).
10. T. M. D. Pereira, *et al.*, *Astrophys. J. Lett.* **792**, L15 (2014).
11. S. R. Cranmer, L. N. Woolsey, *Astrophys. J.* **812**, 71 (2015).
12. H. Iijima, T. Yokoyama, *Astrophys. J.* **848**, 38 (2017).
13. J. Martínez-Sykora, *et al.*, *Science* **356**, 1269 (2017).
14. J. Y. Ding, et al., *Astron. Astrophys.* **535**, A95 (2011).
15. V. Yurchyshyn, V. Abramenko, P. Goode, *Astrophys. J.* **767**, 17 (2013).
16. N. Deng, *et al.*, *Astrophys. J.* **799**, 219 (2015).
17. W. Cao, et al., *Astronomische Nachrichten* **331**, 636 (2010).
18. Materials and methods are available as supplementary materials.
19. W. Cao, et al., *Astronomical Society of the Pacific Conference Series* **463,** 291 (2012).
20. R. L. Moore, A. C. Sterling, J. W. Cirtain, D. A. Falconer, *Astrophys. J. Lett.* **731**, L18 (2011).
21. A. C. Sterling, R. L. Moore, *Astrophys. J. Lett.* **828**, L9 (2016).
22. T. Amari, J.-F. Luciani, J.-J. Aly, *Nature* **522**, 188 (2015).
23. F. Moreno-Insertis, J. Martínez-Sykora, V. H. Hansteen, D. Muñoz, *Astrophys. J. Lett.* **859**, L26 (2018).
24. D. A. Lamb, C. E. DeForest, H. J. Hagenaar, C. E. Parnell, B. T. Welsch, *Astrophys. J.* **674**, 520 (2008).
25. J. R. Lemen, *et al.*, *Sol. Phys.* **275**, 17 (2012).
26. H. Ji, W. Cao, P. R. Goode, *Astrophys. J. Lett.* **750**, L25 (2012).
27. J. A. Klimchuk, *J. Geophy. Res.* **117**, A12102 (2012).
28. M. L. Goodman, *Astrophys. J.* **785**, 87 (2014).

29. H. Peter, P. G. Judge, *Astrophys. J.* **522**, 1148 (1999).
30. S. W. McIntosh, H. Tian, M. Sechler, B. De Pontieu, *Astrophys. J.* **749**, 60 (2012).
31. F. Wöger, O. II von der Lühe, *Proc. SPIE Conf. Ser.* **7019,** 70191E (2008).
32. W. T. Thompson, *Astron. Astrophys.* **449**, 791 (2006).
33. M. DeRosa, G. Slater, Guide to SDO Data Analysis (https://www.lmsal.com/sdodocs/doc/dcur/SDOD0060.zip/zip/entry/).
34. P. Boerner, *et al.*, *Sol. Phys.* **275**, 41 (2012).
35. M. J. Martínez González, L. R. Bellot Rubio, *Astrophys. J.* **700**, 1391 (2009).
36. L. Bellot Rubio, D. Orozco Suárez, *Living Reviews in Solar Physics* **16**, 1 (2019).
37. T. L. Riethmüller, *et al.*, *Astrophys. J. Lett.* **723**, L169 (2010).
38. M. Collados, *Astronomical Society of the Pacific Conference Series* **236,** 255 (2001).
39. J. O. Stenflo, *Astron. Astrophys. Rev.* **21**, 66 (2013).
40. R. Centeno, et al., *Astrophys. J. Lett.* **666**, L137 (2007).
41. B. W. Lites, *et al.*, *Astrophys. J.* **672**, 1237 (2008).
42. S. Şahin, *et al.*, *Astrophys. J.* **873**, 75 (2019).
43. M. C. M. Cheung, M. Schüssler, T. D. Tarbell, A. M. Title, *Astrophys. J.* **687**, 1373 (2008).
44. L. H. Strous, C. Zwaan, *Astrophys. J.* **527**, 435 (1999).
45. R. Centeno, et al., *Astrophys. J. Supp.* **229**, 3 (2017).
46. S. Vargas Domínguez, L. van Driel-Gesztelyi, L. R. Bellot Rubio, *Sol. Phys.* **278**, 99 (2012).
47. P. Gömöry, *et al.*, *Astron. Astrophys.* **511**, A14 (2010).
48. B. Filippov, S. Koutchmy, J. Vilinga, *Astron. Astrophys.* **464**, 1119 (2007).
49. L. Yang, *et al.*, *Astrophys. J.* **852**, 16 (2018).
50. J. J. González-Avilés, *et al.*, *Astrophys. J.* **856**, 176 (2018).
51. T. Kudoh, K. Shibata, *Astrophys. J.* **514**, 493 (1999)
52. P. G. Judge, A. Tritschler, B. Chye Low, *Astrophys. J. Lett.* **730**, L4 (2011).
53. J. Martínez-Sykora, B. De Pontieu, V. Hansteen, S. W. McIntosh, *Astrophys. J.* **732**, 84 (2011).
54. D. H. Sekse, L. Rouppe van der Voort, B. De Pontieu, *Astrophys. J.* **752**, 108 (2012).
55. I. S. Requerey, *et al.*, *Astrophys. J.* **810**, 79 (2015).
56. Y. Suematsu, First Ten Years of Hinode Solar On-Orbit Observatory, eds. T. Shimizu, S. Imada, M. Kubo, *Astrophysics and Space Science Library* **449**, 27 (2018)
57. E. R. Priest, C. E. Parnell, S. F. Martin, *Astrophys. J.* 427, 459 (1994).
58. F. Ellerman, *Astrophys. J.* **46**, 298 (1917).
59. H. Peter, *et al.*, *Science* **346**, 1255726 (2014).
60. J. Wang, Z. Shi, *Sol. Phys.* **143**, 119 (1993).
61. K. Shibata, *et al.*, *Science* **318**, 1591 (2007).

62. L. Ni, Q.-M. Zhang, N. A. Murphy, J. Lin, *Astrophys. J.* **841**, 27 (2017).
63. Y. Shen, Y. Liu, J. Su, Y. Deng, *Astrophys. J.* **745**, 164 (2012).
64. N. E. Raouafi, *et al.*, *Space Sci. Rev.* **201**, 1 (2016).
65. N. K. Panesar, A. C. Sterling, R. L. Moore, *Astrophys. J.* **853**, 189 (2018).
66. T. Yokoyama, K. Shibata, *Nature* **375**, 42 (1995).
67. E. Khomenko, N. Vitas, M. Collados, A. de Vicente, *Astron. Astrophys.* **618**, A87 (2018).
68. E. Khomenko, *Plasma Physics and Controlled Fusion* **59**, 014038 (2017).
69. D. A. Uzdensky, J. C. McKinney, *Physics of Plasmas* **18**, 042105 (2011).
70. A. Brandenburg, E. G. Zweibel, *Astrophys. J. Lett.* **427**, L91 (1994).
71. B. De Pontieu, *et al.*, *Sol. Phys.* **289**, 2733 (2014).
72. L. Rouppe van der Voort, B. De Pontieu, T. M. D. Pereira, M. Carlsson, V. Hansteen, *Astrophys. J. Lett.* **799**, L3 (2015).
73. B. De Pontieu, J. Martínez-Sykora, G. Chintzoglou, *Astrophys. J. Lett.* **849**, L7 (2017).
74. M. S. Madjarska, K. Vanninathan, J. G. Doyle, *Astron. Astrophys.* **532**, L1 (2011).
75. Y. Z. Zhang, *et al.*, *Astrophys. J.* **750**, 16 (2012).
76. F. Jiao, et al., *Astrophys. J. Lett.* **809**, L17 (2015).
77. T. Samanta, V. Pant, D. Banerjee, *Astrophys. J. Lett.* **815**, L16 (2015).
78. V. M. J. Henriques, D. Kuridze, M. Mathioudakis, F. P. Keenan, *Astrophys. J.* **820**, 124 (2016).
79. E. R. Priest, J. F. Heyvaerts, A. M. Title, *Astrophys. J.* **576**, 533 (2002).
80. K. J. Li, J. C. Xu, W. Feng, *Astrophys. J. Suppl. Ser.* **237**, 7 (2018).
81. D. W. Longcope, L. A. Tarr, *Phil. Trans. R. Soc. A* **373**, 20140263 (2015).
82. J. Zhang, et al., *Astrophys. J. Lett.* **799**, L27 (2015).
83. N.-E. Raouafi, G. Stenborg, *Astrophys. J.* **787**, 118 (2014).
84. L. P. Chitta, et al., *Astrophys. J. Supp.* **229**, 4 (2017).
85. E. R. Priest, L. P. Chitta, P. Syntelis, *Astrophys. J. Lett.* **862**, L24 (2018).
86. P. Syntelis, E. R. Priest, L. P. Chitta, *Astrophys. J.* **872**, 32 (2019).
87. H. Tian, S. W. McIntosh, B. De Pontieu, *Astrophys. J. Lett.* **727**, L37 (2011).
88. B. De Pontieu, S. W. McIntosh, *Astrophys. J.* **722**, 1013 (2010).

**References (31-88) are called out only in the Supplementary Materials (SM).**

**Acknowledgments:** BBSO operation is supported by NJIT and NSF. GST operation is partly supported by the Korea Astronomy and Space Science Institute (KASI); Seoul National University; the Strategic Priority Research Program of CAS (grant no. XDB09000000); and the Operation, Maintenance and Upgrading Fund for Astronomical Telescopes and Facility Instruments administrated by CAS. The AIA is an instrument on SDO, a mission of NASA's Living With a Star Program. We thank the GST and SDO teams for providing the data and S. K. Solanki and L. R. Bellot Rubio for helpful discussion.
**Funding:** This work is supported by NSFC grants 11825301, 11790304(11790300), 41574166, 11729301, U1931107, and 11850410435; NSF grants AGS 1821294, 1620875, and AST-1614457;

AFOSR grant FA9550-19-1-0040, NASA grants HGC 80NSSC17K0016 and HGI 17-HGISUN17_2-0047; Max Planck Partner Group program, Strategic Priority Research Program of CAS (grant XDA17040507), Key Applied Basic Research Program of Yunnan Province (FS: 2018FA035), FWF project N27800, CAS Presidents International Fellowship Initiative grant no. 2019VMA052, UK STFC grant ST/M000826/ 1, and the Royal Society. **Author contributions:** H.T. and T.S. conceived the study and wrote the manuscript. T.S. analyzed the data and generated the figures and movies under H.T.'s guidance. V.Y. generated the GST observations, processed the GST data, and advised on the data analysis. W.C. developed instruments on GST. K.A. processed the GST NIRIS data for scientific use. S.F. helped co-align the data. Y.C. performed the energy calculation and helped verify the results. H.P., A.S., R.E., D.U., and D.B. contributed to the interpretation of the observations. All authors discussed the results and commented on the manuscript. **Competing interests:** There are no competing interests. **Data and materials availability:** The GST dataset that we used is available at http://ftp.bbso.njit.edu/pub/20170619/. The AIA data are available at the Joint Science Operations Center, http://jsoc.stanford.edu/AIA/ AIA_lev1.html; we used the 12 s 171 Å and 24 s 1700 Å observations in the time range 2017 June 19 18:45-18:57 UT.

# Supplementary Materials for

# Generation of Solar Spicules and Subsequent Atmospheric Heating

Tanmoy Samanta, Hui Tian*, Vasyl Yurchyshyn, Hardi Peter, Wenda Cao, Alphonse Sterling, Robertus Erdélyi, Kwangsu Ahn, Song Feng, Dominik Utz, Dipankar Banerjee, Yajie Chen

*Correspondence to: huitian@pku.edu.cn

**This PDF file includes:**
    Materials and Methods
    Supplementary Text
    Figs. S1 to S10
    Table S1
    Captions for Movies S1 to S9

**Other Supplementary Materials for this manuscript include the following:**
    Movies S1 to S9

# 1. Materials and Methods

## 1.1 Observations

### 1.1.1 Goode Solar Telescope

We used data acquired by the Broad-band Filter Imager (BFI), the Visible Imaging Spectrometer (VIS) and the Near Infra-Red Imaging Spectropolarimeter (NIRIS; *19*) of the 1.6-m Goode Solar Telescope (GST; *17*) at the Big Bear Solar Observatory. The BFI instrument obtains continuum images with a broad-band filter centered at the head of the photospheric molecular titanium oxide band (TiO 7057 Å, 10-Å bandpass). The VIS instrument provides narrow-band Hα line-scan images. The NIRIS instrument obtains magnetograms of the photosphere through spectropolarimetric measurements of the Fe I 1.56 μm infrared line.

During our observations on 2017 June 19, the TiO filtergrams were taken in bursts of 100 frames. The VIS instrument uses a single Fabry-Pérot etalon to produce a narrow 0.07-Å bandpass within a range of 550-700 nm. Bursts of 25 frames were taken at the blue and red wings of the Hα line (± 0.8 Å) to achieve a high cadence. All bursts were flat-field corrected, and processed using speckle reconstruction (*31*) to achieve diffraction-limited resolutions. Afterwards, they were aligned, and de-stretched to remove residual image distortion due to seeing and telescope jitter. Examples of the Hα line wing images are shown in Fig. S1, where spicules are seen as the elongated dark features and form a rosette structure (*1*). NIRIS utilizes dual Fabry–Pérot etalons to perform spectropolarimetric measurements at 58 wavelength positions (36 were used in our analysis) using the 1.56 μm line. The step of the spectral scan was 0.11 Å. Each frame was flat-field corrected, co-aligned and de-stretched. Polarization calibration was applied to the NIRIS data to remove cross-talk (*17, 19*). After calibration, the root-mean-square (rms) noises in the spectral continua of Stokes-*Q*, *U,* and *V* are found to be 0.08-0.13% (for different frames) of the Stokes-*I* continuum intensity. The Fe I 1.56 μm line has an effective Landé factor of 3, and its near-infrared wavelength provides an advantage compared to optical lines in measuring the weak magnetic fields in internetwork regions (regions surrounded by the magnetic network). A summary of the observations, including the pixel sizes and cadences of our reconstructed image sequences, are listed in Table S1. The center of our region of interest (ROI) shown in Fig. S1B-E is given in Helioprojective-Cartesian coordinate system (*32*). The diffraction limited resolution of the telescope with adaptive optics is 46 km at 5000 Å, and 146 km at 1.56 μm.

### 1.1.2 Solar Dynamics Observatory

We also used observations from the AIA telescope on the SDO spacecraft. The AIA images were calibrated using the standard *aia_prep.pro* routine (*33*) in SolarSoft (SSW). The resolution and cadence of the AIA data are listed in Table S1. The field-of-view (FOV) of our observed region on the Sun is indicated in Fig. S1A.

The AIA and GST data were co-aligned by matching the locations of network bright points observed in both the AIA 1700 Å and GST TiO images.

The AIA 171 Å images were utilized to study the coronal response of the chromospheric spicules. The temperature response function of this passband peaks at a temperature of ~0.8 MK (*34*).

## 1.2 Methods

**1.2.1 Derivation of the line-of-sight magnetic fields and Dopplergrams**

We used the calibrated NIRIS data to measure the photospheric magnetic fields. Because the Stokes-$Q$ and Stokes-$U$ signals are generally weak in the quiet-Sun region, we mainly focus on the Stokes-$I$ and Stokes-$V$ profiles, which provide information on the line-of-sight (LOS) magnetic fields. Because our observed region is close to the disk center, the LOS magnetic fields are approximately the longitudinal component of the magnetic fields ($B_L$). We used the weak-field approximation (WFA) to determine the longitudinal flux density (e.g., *35, 36*). An analytical least-squares model fitting was applied to reduce the dependence on noise (e.g., *35, 36*). The relationship between the longitudinal flux density ($\phi_L$) and the Stokes-$I$/$V$ spectra can be expressed as

$$\phi_L = fB_L = -\frac{\sum_i \frac{dI}{d\lambda_i} V_i}{4.6686 \times 10^{-13} \lambda_0^2 g_{\text{eff}} \times \sum_i \left(\frac{dI}{d\lambda_i}\right)^2} \quad (S1)$$

where $\lambda_0$ represents the rest wavelength, $g_{\text{eff}}$ is the effective Landé factor of the Fe I 1.56 μm line (equals 3), and the index $i$ runs through the different wavelength positions. Here both $\lambda_0$ and $\lambda_i$ are expressed in Å. The filling factor $f$ accounts for the scenario that the field occupies only a fraction of the pixel area. Because the filling factor is unknown, we just estimated the longitudinal flux density from the right side of Equation S1. One image of the longitudinal flux density obtained from this method is shown in Fig. S2A.

We selected all pixels where the longitudinal flux density derived from the WFA is larger than 10 Mx cm$^{-2}$, then examined the Stokes profiles at these pixels and excluded pixels where the polarization signal is likely caused by noise. For this purpose, we fitted a single Gaussian model to the Stokes-$I$ profile to estimate the continuum intensity ($I_{\text{cont}}$) at each pixel. From this fitting, we also determined the centroid of the line (e.g., *37*), and computed the Doppler velocity at each pixel by assuming that the average Doppler velocity in the full FOV is zero. Afterwards, we normalized the Stokes-$V$ profile to the Stokes-$I$ continuum intensity. Then, we fitted a two-Gaussian function with opposite amplitudes to the Stokes-$V/I_{\text{cont}}$ profile around the line center. Two examples of the fitted profiles are shown in Fig. S2B and C. We then picked out the pixels where the fitted peak is higher than three times the standard deviation of the residual profile (3σ). We have also visually inspected many fitted profiles to check whether they are well-reproduced by the double Gaussian model. Our visual inspection suggests that many pixels with fitted peaks smaller than 3σ do not have adequate signals, and we have excluded those pixels from the magnetograms. We have also excluded many individual pixels that appear to be random and isolated. Detected magnetic elements that are less than 3 pixels in size were also excluded.

After this, we plotted the peak value of the fitted Stokes-$V/I_{\text{cont}}$ profile ($V/I_{\text{cont}}$ in short) against the longitudinal flux density (Fig. S2D). The plot shows that the $V/I_{\text{cont}}$ and the longitudinal flux density have a nearly linear relationship in the weak-field regime (up to ~200 Mx cm$^{-2}$). From this linear behavior, we found that a $V/I_{\text{cont}}$ value of 0.35% is roughly equivalent to a flux density of 10 Mx cm$^{-2}$. At some pixels the $V/I_{\text{cont}}$ is much higher than 0.35% and the flux density is lower than 10 Mx cm$^{-2}$, and vice-versa at some other pixels. We visually inspected the model fitting at those pixels, and found that the signals are usually weak and that the fitting may not be reliable. So we have further excluded those pixels from the magnetograms.

We then applied a binary mask to the magnetograms, which preserves the information on the direction of the magnetic fields (positive or negative). Figure S2E presents a binary magnetic field map which shows only magnetic elements with a longitudinal flux density of at least 10 Mx cm$^{-2}$, a $V/I_{cont}$ signal of at least 0.35% and larger than 3σ, and at least 3 pixels in size. The Stokes-$V$ profiles are generally antisymmetric at these pixels. Examples of the binary magnetic field maps are also presented in Figs. 1-3, Figs. S4, S6 & S7.

The Fe I 1.56 μm line is a highly magnetic-sensitive line, and the WFA breaks down above ~500 Gauss (*38, 39*). We are mainly interested in the weak opposite-polarity (negative-polarity in our observations, opposite to the dominant polarity in the magnetic network) magnetic fields around the strong kilo-Gauss network fields. In our observations the flux density of the opposite-polarity fields generally ranges from ~10 to 100 Mx cm$^{-2}$ (Gauss), for which the WFA is valid. The WFA is likely not a good approximation for the network fields with a strength higher than ~500 Gauss. However, even in that case, the directions of the magnetic fields can be reliably derived. So the presented binary magnetic field maps, which only have the information about the directions of longitudinal magnetic fields, are not affected by the WFA.

**1.2.2 Calculation of the fractional net linear polarization**

We focus on the interaction between weak internetwork fields and strong network fields. The magnetic fields of the quiet-Sun regions, especially the internetwork fields, are much weaker than those of active regions. Hence, investigating the properties of these internetwork fields, particularly the horizontal fields, is difficult since the linear polarization signals (Stokes-$Q$ and $U$) are very weak. However, the nearly horizontal fields associated with the apexes of emerging magnetic bipoles may result in detectable linear polarization in internetwork regions (*40, 41*). To search for signatures of horizontal fields from the NIRIS data, we computed the fractional net linear polarization ($L_{tot}$; *41*) defined as

$$L_{tot} = \frac{\sum_i [Q_i^2 + U_i^2]^{1/2} d\lambda_i}{I_{cont} \sum_i d\lambda_i} \qquad (S2)$$

The integration was carried out from 15648.0 Å to 15649.0 Å. A snapshot of the fractional net linear polarization obtained using Equation S2 is shown in Fig. S3B. Several small patches of strong linear polarization can be seen in our FOV. Some of these linear polarization signals are associated with flux emergence (Figs. S4 & S5).

# 2. Supplementary Text

## 2.1 Flux emergence/appearance

Previous studies show that observationally there are two types of flux emergence in the internetwork region: emergence of bipolar and unipolar elements. The latter is sometimes called flux appearance, and is far more common than the former (*36*). A patch of unipolar flux should be connected to a flux of opposite polarity. However, in most cases only one polarity is observed, and no obvious opposite polarity is found near the emergence site (*24*). The physical mechanism responsible for the appearance of unipolar flux is poorly understood. Observations have shown that one polarity in an emerging bipole is often more compact and stronger than the other one (e.g., *42*). So it is likely that the magnetic flux at the opposite-polarity footpoint of an emerging loop is more dispersed and thus below the detection limit of the instrument. However, it is also possible that the appearance of a unipolar patch results from existing fluxes which coalesce and become strong enough to be detectable (*24*). These previous observations have shown that the emergence of a typical bipole is not common, and that the emergence/appearance of unipolar elements is much more common in the internetwork region. Our observations are consistent with these findings.

Our conclusion that spicules can be generated by magnetic reconnection between the dominant network fields and small-scale opposite-polarity fields does not necessarily require the presence of a bipole. The presence of a small-scale opposite-polarity magnetic element at the spicule footpoint, which we observe, is sufficient to support this mechanism. As mentioned above, a small-scale opposite-polarity magnetic element could be one footpoint of an emerging loop or result from coalescence of existing elements.

Nevertheless, we do find signatures of bipole emergence leading to the generation of a few spicules. One example is the event shown in Fig. 2A-D. We show our analysis of the polarization signals for this event in Figs. S4 & S5. From Fig. S4, we first see the appearance of a small patch of linear polarization above a granule at t=6.50 min, which likely corresponds to the apex of an emerging magnetic loop (e.g., *40*). In the meantime, we see elongation of the granulation and a transient darkening, also consistent with the scenario of granule-scale flux emergence (*43, 44*). The linear polarization patch becomes larger as the bipole rises, and about one minute later we see the appearance of a patch of negative-polarity longitudinal flux between the linear polarization patch and the strong network flux. At t=9.32 min, the linear polarization signal disappears, suggesting that the loop apex has risen to a higher layer. At the same time, the area of the longitudinal flux patch increases. This patch should be one footpoint of the rising loop. The circular polarization signal at the other footpoint appears to be very weak. However, by adding the Stokes-*V* signals over a large area at the expected location of the other footpoint, we detect the other polarity (Fig. S5). Figure S4 also shows the appearance of an enhanced spicular activity at t=9.32 min, when the linear polarization signal disappears. This spatiotemporal correlation strongly suggests the generation of the enhanced spicular activity by magnetic reconnection between the rising magnetic loop and the overlying network fields. We noticed that a positive-polarity patch (just to the right of the rectangle in Fig. S4A) is present at almost the same time. This might be a coincidence, because the enhanced spicular activity appears to be initiated from the emergence site of the negative-polarity flux rather than the site of this positive-polarity flux (Movie S2).

The flow pattern is expected to be complicated for granule-scale flux emergence. The classical scenario of flux emergence in the form of a rising loop includes upflow and downflows at the loop apex and footpoints, respectively (e.g., *45*). However, convective flows, draining along loop legs and upward motion of the entire loop all happen at the same time, leading to a complex flow pattern that is often different from the classical one. In many cases, blueshifts are found in photospheric lines at both the loop apexes and footpoints, and draining downflows are seen in low-chromospheric lines at the loop footpoints (e.g., *36*). In some cases, photospheric upflows are more prominent at the loop footpoints (e.g, *35, 46*). However, sometimes the scenario is different and more complicated. Chromospheric downflows could appear in one or both footpoints during the emergence, without obvious downflows in the photosphere (e.g., *35*). While for some other cases, downflows are seen in photospheric lines during the later stage of flux emergence (e.g., *47*). These previous observations suggest that we should not expect to see a simple flow pattern for small-scale flux emergence events. Nevertheless, the sudden enhancement of blue shift at the location of the opposite polarity in the event shown in Fig. 2D agrees with previous observations of flux emergence mentioned above.

## 2.2 Formation mechanism of spicules

At any given instant, around one million spicules eject plasma from the Sun's surface (*1*). There exist many proposals for the formation mechanisms of spicules, such as magneto-acoustic shocks (e.g., *3*) or magnetic reconnection (e.g., *14, 48–50*). It has also been suggested that spicules may be generated by an upward force due to Alfvén waves (e.g., *11, 12, 51*). A different suggestion is that the high-speed spicules could be warps in two-dimensional sheet-like structures (*52*). Through numerical simulations, it was suggested that large field gradients and intense electric currents squeeze the dense chromospheric material, which produces heating and accelerates spicules (*53*). Recent 2.5D numerical simulations (*13*) have produced numerous spicules by magnetic tension that is amplified and transported upward through ambipolar diffusion. The initial buildup of strong magnetic tension directly results from the interaction of magnetic flux concentrations in the Sun's network region with the constantly emerging weak and horizontal fields.

Direct observational evidence for these ideas are generally missing. One observational study (*16*) examined the possible connection of chromospheric upflow events to flux emergence and/or cancellation near network boundaries. Their analysis appears to reveal no obvious correlation. We noticed that most of these upflow events have a round or elliptical shape, which is different from the needle-like morphology of spicules in higher-resolution (spatial, temporal and spectral) observations (e.g., *54*). These events might correspond to certain segments of the spicules, or they might not represent the type of spicules we have observed in our data. It is also possible that the lower resolution and lower sensitivity of the earlier magnetic field measurements have affected the detection of weak opposite-polarity magnetic elements at the footpoints of some spicules.

Evidence for the formation mechanism of spicules can be provided through simultaneous, high-cadence and high-resolution observations of both the photospheric magnetic fields and chromospheric dynamics. The 1.6-m GST telescope is well suited for such observations. Our high-resolution Hα imaging observations allow us to identify the footpoint locations of spicules and investigate the magnetic field evolution right there. In addition, our magnetograms were obtained with the Fe I 1.56 μm line formed in the deep photosphere, where the magnetic fields of the opposite-polarity elements are relatively strong. Its long wavelength and large effective Landé factor also help to measure the weak magnetic elements. Because of these key differences, we find evidence for spicule generation by the interaction between magnetic fields with different polarities.

In our observations, we identified 23 enhanced spicular activities. Except for one event that was already ongoing at the beginning of our observations, the other 22 events all occur in conjunction with the co-spatial and co-temporal presence of negative-polarity magnetic elements around their footpoints near the network boundary. The footpoint locations of all these enhanced spicular activities are marked by the black circles in Movie S4. Around two thirds of the enhanced spicular activities are found during the flux emergence/appearance stage, and some of them are followed by flux cancellation. The other third occur during the stage of flux cancellation. The enhanced spicular activities mostly occur in the East and North parts of the FOV. The southwest part shows reduced magnetic fields and fewer photospheric bright points (proxies of network magnetic elements), and consequently the spicular activity is much lower in this region compared to the other parts of the network (see Movie S4). There is only one enhanced spicular activity in this region, which appears to be related to the interaction of a weak negative-polarity flux with a patch of strong positive-polarity flux (Fig. S3, Movie S4).

As mentioned in the main text, a plausible scenario consistent with our findings is that spicules are generated by magnetic reconnection between weak opposite-polarity fields and the expanded network fields in the chromosphere. To generate spicules the opposite-polarity magnetic structures may need to be close to the network and rise to higher layers (possibly to the chromosphere). In our observations, about two thirds of the negative-polarity magnetic elements within the outer white circle (3 Mm from the approximate network edge) in Fig. 3A are associated with the generation of one or more spicules during their lifetimes. These magnetic elements are found at the footpoints of spicules, similar to the results presented in Figs. 1, 3 and S7. In a few of these cases, we observed enhanced spicular activities ~1–5 minutes after the first appearance of the opposite-polarity elements (e.g., Fig. S4). These results agree with the scenario described above. The remaining one third (mostly small-size) do not produce any noticeable spicules during their lifetimes, and some of them may represent low-lying elements (*35*) that fail to interact with the overlying network fields. Compared to the enhanced spicular activities, some individual spicules are related to a smaller patch of the negative polarity at their footpoints. Thus, it is possible that very small or weak magnetic elements lead to very small-scale spicules that cannot be resolved with GST's resolution and sensitivity, which may explain the fact that several very small opposite-polarity patches disappear very fast (lasting for 1-2 frames) without noticeable spicules from those locations. So it appears that the distance to the network, the height the opposite-polarity magnetic structure reaches, and the magnitude of magnetic flux possibly all play a role in the generation of spicules.

Network fields often contain bundles of individual flux tubes (e.g., *55*). Component reconnection (reconnection between nonantiparallel magnetic fields) between these differently oriented flux tubes inside the unipolar network has been proposed as a possible mechanism for spicules generation (*56*). Our observations of opposite-polarity magnetic elements around spicule footpoints do not support this scenario. However, we cannot exclude the possibility that some of the faint individual spicules are generated by this process.

To understand the energetics of spicules, we have performed an order-of-magnitude estimation of the kinetic energy of spicules and the available magnetic energy around the feet of spicules. From our observations, we found a typical width ($d$) of 200 km, a length ($L$) of 3,000 km and a speed ($v$) of 50 $km\ s^{-1}$ for individual spicules. Taking a density ($\rho$) of $10^{-13} g\ cm^{-3}$ (e.g., *1, 2*), we found that the kinetic energy ($E_k$) of a spicule is

$$E_k = \frac{1}{2}\rho \cdot \pi \frac{d^2}{4} L \cdot v^2 \approx 1.2 \times 10^{23}\ erg \quad (S3)$$

The magnetic field strength ($B$) and area ($A$) of an opposite-polarity magnetic element are around 20 G (assuming a filling factor of unity) and 0.75 $Mm^2$, respectively. Assuming a loop length ($l$) of 2000 km, the magnetic energy ($E_B$) can be calculated as

$$E_B = \frac{B^2}{2\mu_0} \cdot l \cdot A \approx 2.5 \times 10^{25}\ erg \quad (S4)$$

This calculation suggests that the available magnetic energy is sufficient to drive spicules. The enhanced spicular activities generally consist of groups of individual spicules, and the field strengths and sizes of the opposite-polarity patches at their footpoints are also larger (about 30 G and 2.5 $Mm^2$, respectively). As a result, the kinetic energy and available magnetic energy are both ~1–2 orders of magnitude larger than those of an individual spicule.

The scenario of magnetic reconnection between one pole of an existing/emerging magnetic loop and the nearby network fields, as mentioned above, suggests that the disappeared photospheric flux could be an approximate measure of the flux cancelled through reconnection. From our photospheric magnetic field measurements, we found that the cancelled flux is of the order of $10^{16} Mx$ for a typical spicule. Considering a typical cancellation period of ~3 minutes, the flux cancellation rate should be around $5 \times 10^{13} Mx\ s^{-1}$. Since at any moment there are around one million *(1)* spicules on the Sun, the flux cancellation rate should be around $5 \times 10^{19} Mx\ s^{-1}$ to generate all these spicules.

In principle the observed flux cancellation/disappearance in the photosphere could also be caused by other processes such as flux dispersal, flux dissipation, emergence of U-shaped loops, or submergence of Ω-shaped loops. However, the observed correlation between flux cancellation/disappearance and spicule generation suggests that a large fraction of the observed flux cancellation/disappearance within the outer white circle of Fig. 3A should be a signature of a certain physical process that produces the spicules. There are no existing suggestions that the above-mentioned mechanisms could lead to the generation of spicules in the absence of interaction with the network fields. On the other hand, reconnection between weak internetwork fields and the strong network fields, which can lead to the observed signals of flux cancellation/disappearance, is consistent with some proposed spicule-generation mechanisms (e.g., *14*). Hence, it is likely that the cancellation/disappearance events responsible for the generation of spicules are signatures of reconnection. In addition, submergence of Ω-loops often naturally results from reconnection at higher layers (e.g., *57*), which is consistent with our scenario. U-loops may form during flux emergence. However, these U-loops are normally observed in active regions, and reconnection in these U-loops is usually related to the generation of Ellerman bombs/Ultraviolet bursts (e.g., *58, 59*) rather than spicules. As we mentioned above, some of the opposite-polarity magnetic elements do not produce any noticeable spicules. Part of them might represent low-lying magnetic field structures *(35)* that fail to interact with the nearby network fields, and their disappearance could be due to the other processes mentioned above instead of reconnection between internetwork and network fields.

Small-scale opposite-polarity magnetic elements are also found at the footpoints of many individual spicules (e.g., Figs. 3 & S7), suggesting the possibility of the enhanced spicular activities and small individual spicules being produced by the same reconnection mechanism mentioned above. Theoretically it is poorly understood why some spicules appear as enhanced spicular activities while others appear as isolated individual spicules. However, from an observational point of view, the simultaneous generation of many spicules in an enhanced spicular activity might be related to two factors. First, the larger available magnetic energy could be sufficient to drive more than one spicule. Second, the larger size of the opposite-polarity element means a larger area of interaction with the network fields, favoring the occurrence of reconnection at multiple sites within a wider region. Due to the observational limitations, we cannot exclude the possibility that some individual spicules are produced by other processes.

Our observations of flux emergence leading to the generation of spicules might be consistent with a recent numerical simulation (*13*). However, some of our examples clearly show that flux cancellation (*60*) leads to the formation of spicules. This scenario is not clear from their simulation. Our observations of spicule generation by magnetic reconnection might be in line with various reconnection-driven jet scenarios or models (e.g., *21, 61, 62*), suggesting that spicules may not be that different from these reconnection-driven jets including large coronal jets (e.g., *63-65*). Observations of these jet-like features appear to suggest a continuous spectrum of jets in terms of the size and associated (cancelled) magnetic flux. The different spatial scales of these jets are likely related to the sizes of the reconnected magnetic structures as well as the reconnection heights (*14, 61*). Though our observations indicate that the spicules observed here are driven by magnetic reconnection, they are not necessarily ejected directly from the neutral point of the reconnection. The whip-like motion of the cool plasma in some spicules could also result from the sling-shot effect due to the reconnection (*66*). In this case, there could be a spatial offset between the reconnection site and the exact footpoint of a spicule.

We expect that the reconnection process responsible for the generation of spicules occurs in the lower solar atmosphere, where the plasma is partially ionized. The existence of a large number of neutrals leads to the decoupling of the plasma and magnetic fields, and lifts the magnetic fields into the higher atmosphere easily through the ambipolar diffusion process (*67*). These magnetic field structures then interact with the expanded network fields possibly in the chromosphere. Ambipolar diffusion could be an important process in the generation of spicules (*13*), as it can bring the magnetic field lines sufficiently close to facilitate reconnection (*68*). Ambipolar diffusion, as well as radiative cooling, can increase the reconnection rate because both processes can result in a strong current sheet thinning and current density enhancement (*69, 70*).

## 2.3 Coronal connection of spicules

Spicules may serve as an energy/mass bridge between the cool photosphere and the million-degree corona. Based on Interface Region Imaging Spectrograph (IRIS; *71*) observations, it was suggested that some spicules are heated to at least ~100,000 K, appearing as fast intermittent network jets in transition region images (*9*). Such a suggestion has been confirmed through multi-instrument observations (*10, 72*). The fast apparent motions of network jets are likely related to the rapidly moving fronts of spicular heating (*73*). It is under debate whether spicules play a dominant role in coronal heating (e.g., *4, 74, 75*). Observations have shown signatures of some chromospheric upflow events connecting to the upward propagating disturbances (PDs) seen in AIA coronal images (e.g., *4, 76, 77*). This connection was mostly observed in active regions and coronal holes. It has been found that low coronal brightenings are rarely associated with spicules (only 6% match with signatures in AIA 171 Å) in a quiet-Sun region (*78*). It is not clear why some spicules show signatures in the corona whereas others do not. Moreover, it is still unclear whether the PDs are plasma flows or slow-mode magneto-acoustic waves, which makes it difficult to evaluate their contribution to coronal heating.

Some analytical models have indicated that the dissipation of electric currents resulting from the shuffling of ubiquitous mixed-polarity fields on small scales could provide coronal heating at low heights (*79*). Observations reveal that the weak magnetic activity in the photosphere effectively correlates with the coronal emission (*80*). Through modeling (*81*) and observations (*82*), it is also

T. Samanta, H. Tian, V. Yurchyshyn, et al., Science 366, 890 (2019)shown that the emergence of new large-scale flux interacts with the overlying magnetic fields and effectively heats the corona. It has also been suggested that magnetic flux cancellation events at the footpoints of coronal plumes or loops are a source of mass and energy supply to the corona (*83-86*). Despite these intensive investigations, it is still not clear how the photosphere and corona are coupled on small scales. It is also unknown how the interaction of magnetic fluxes in the lower atmosphere transfers energy into the corona.Our observations from the photosphere to the chromosphere and corona show that many spicules are driven by the interaction between network fields and opposite-polarity internetwork fields, and that about 20 of the 23 identified enhanced spicular features (with a size of several AIA pixels) during our observing period channel hot plasma into the corona (e.g., Figs. 4, S8, S9, Movie S5). The hot plasma generally shows an apparent upward motion with a velocity of 20-50 km s$^{-1}$ in the plane of sky. We observed a quiet-Sun region. Such regions occupy a much larger area of the solar surface compared to active regions and coronal holes. Thus, our result demonstrates the role of prevalent chromospheric jets in coronal heating under the much more common quiet-Sun condition.

To understand the energy budget of the coronal counterpart of spicules, we have also performed an order-of-magnitude estimation of the coronal thermal energy *($E_T$)*,

$$E_T = \frac{3}{2} n k_B T \cdot V \qquad (S5)$$

Here, *n*, *$k_B$*, *T* and *V* are the coronal electron number density, Boltzmann constant, temperature and volume of the coronal counterpart of a spicule. Again, considering a typical width (*d*) of 200 km and a length (*L*) of 3000 km, we found that the volume ($V = \pi \frac{d^2}{4} L$) of an individual spicule is ~$10^{23}$ $cm^3$. Taking a typical coronal electron density $10^9$ cm$^{-3}$, the thermal energy required to heat a substantial fraction of the spicule to ~1 MK is ~$10^{22}$ $erg$. For the enhanced spicular activities, ~1–2 order of magnitude higher thermal energy is required to cause the same heating. The available magnetic energy estimated from Equation S4 appears to be sufficient to heat both individual spicules and enhanced spicular activities to typical coronal temperatures. This leads to the question of why enhanced spicular activities often show coronal counterparts, whereas small individual spicules do not. The fact that most enhanced spicular activities show coronal signatures implies that the size/activity level is a key factor to determine the coronal response. It is possible that coronal heating events caused by spicules are very common, but their signals associated with small individual spicules are hidden in the noise or unresolved by AIA. However, we cannot rule out the possibility that some spicules are subject to less heating and thus cannot reach coronal temperatures.

Our observations show that the coronal counterparts generally appear at the top of the spicules (Figs. 4, S8 & S9) and that some of these materials fall back from the coronal height following a parabolic path (Fig. S10). The downward motion appears to be present for roughly one fourth of the enhanced spicular activities, and the plane-of-sky velocity is generally in the range of 10-30 km s$^{-1}$. These return flows might be responsible for the prevalent redshift of emission lines formed in the chromosphere-corona transition region (*29, 30*). This also suggests that at least some of the PDs observed in the AIA coronal channels are real mass flows (*87, 88*), rather than slow magneto-acoustic waves. These downflows were rarely reported from previous imaging observations, and their observation completes the poorly understood mass cycle between the chromosphere and corona (*30*).

# 3. Figures

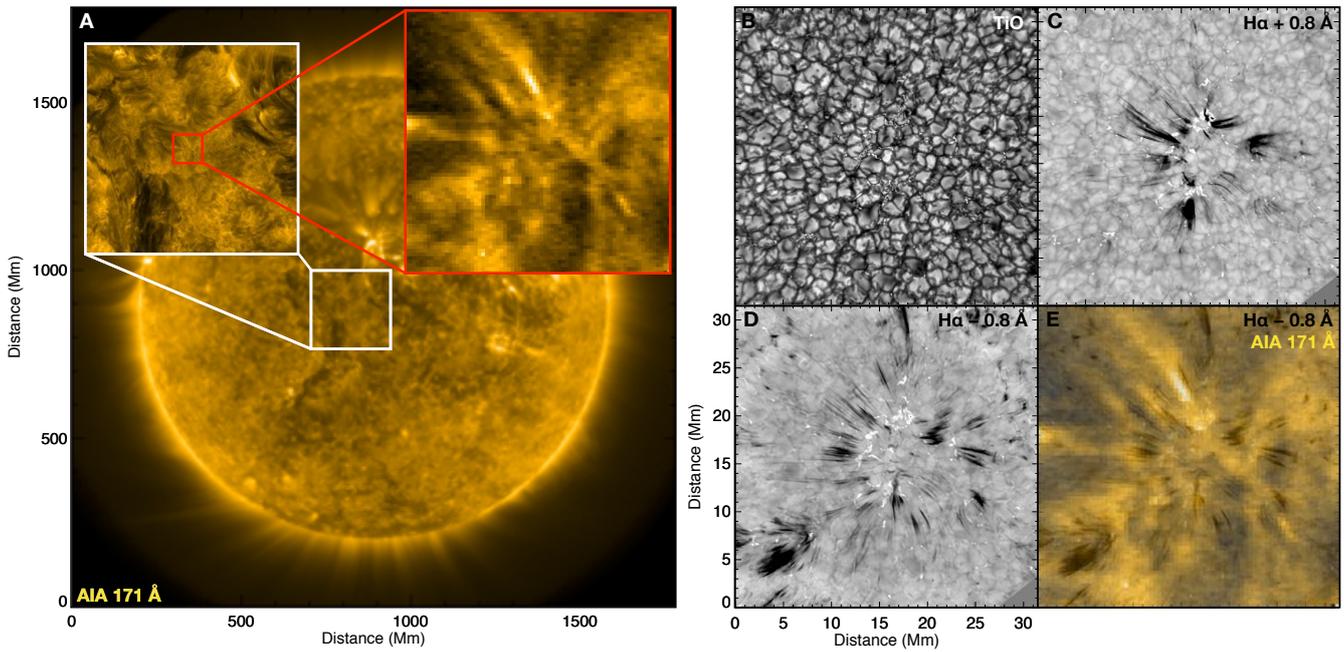

**Fig. S1. AIA and GST images taken around 18:55 UT on 19 June 2017.** (A) Full-disk image of AIA 171 Å. The white inset shows a zoomed in view, whilst the red inset is our region of interest. (B-E) Images at the wavelengths of TiO, Hα red wings, Hα blue wings, and Hα blue wing overlain with AIA 171 Å in the ROI. Movie S6 shows an animated version of panels (B-E).

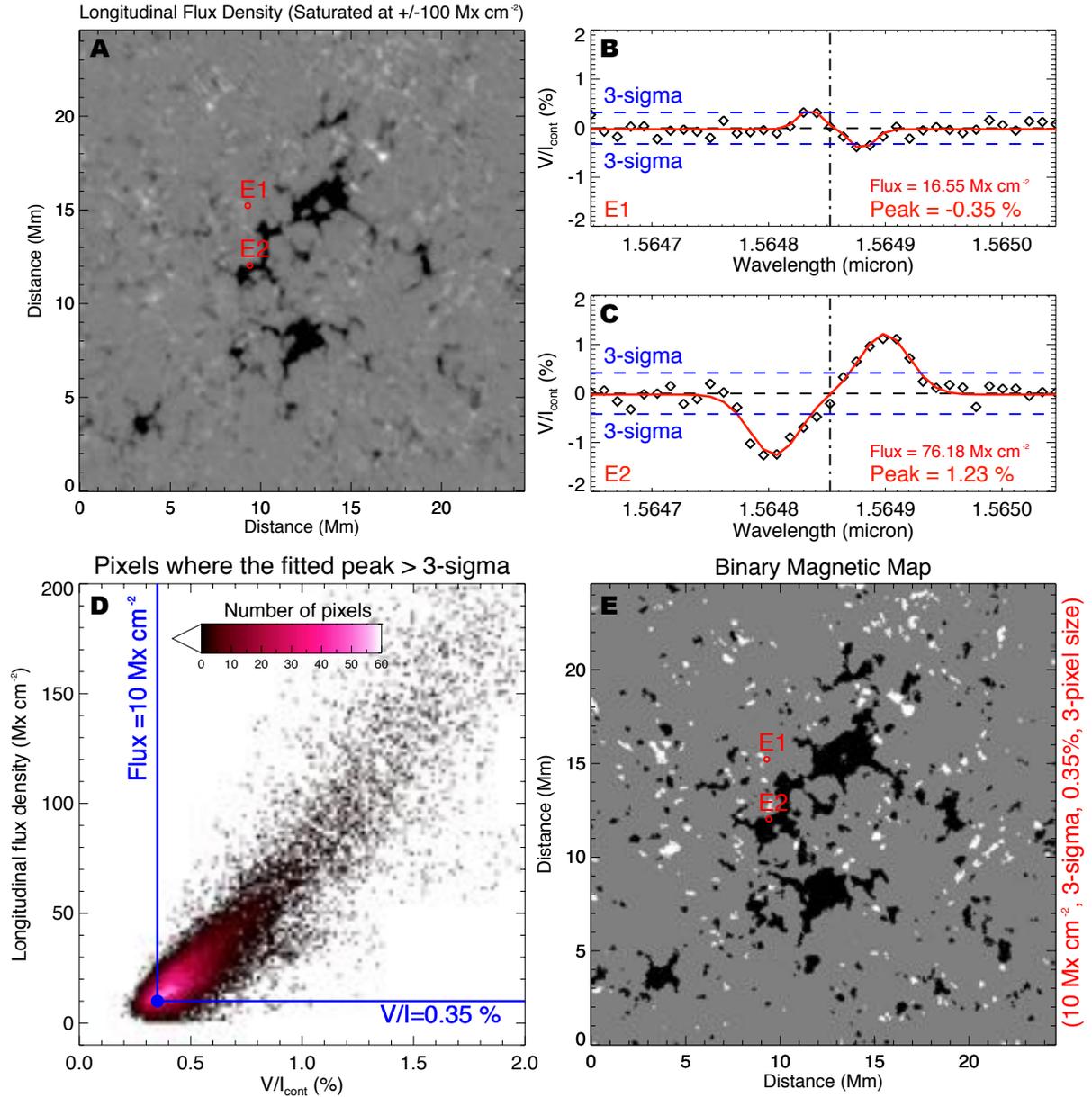

**Fig. S2. Determination of the binary magnetic field maps.** (A) The longitudinal flux density (3-pixel smoothed) in the ROI at 18:46 UT. Negative and positive polarities are shown in white and black, respectively. (B-C) Stokes $V/I_{cont}$ profiles at the locations E1 and E2 marked in panel (A). The red lines show the results of fitting with a two-Gaussian function. The vertical lines indicate the rest wavelength. The horizontal lines mark the levels of zero and +/-3σ (3-sigma). (D) Scatter plot showing the relationship between the longitudinal magnetic flux density estimated from the WFA and the peak value of the fitted $V/I_{cont}$ profile. The number of pixels is represented by a color spectrum in the 2D histogram. Pixels with a fitted peak smaller than 3σ are excluded. (E) The resulting binary magnetic field map.

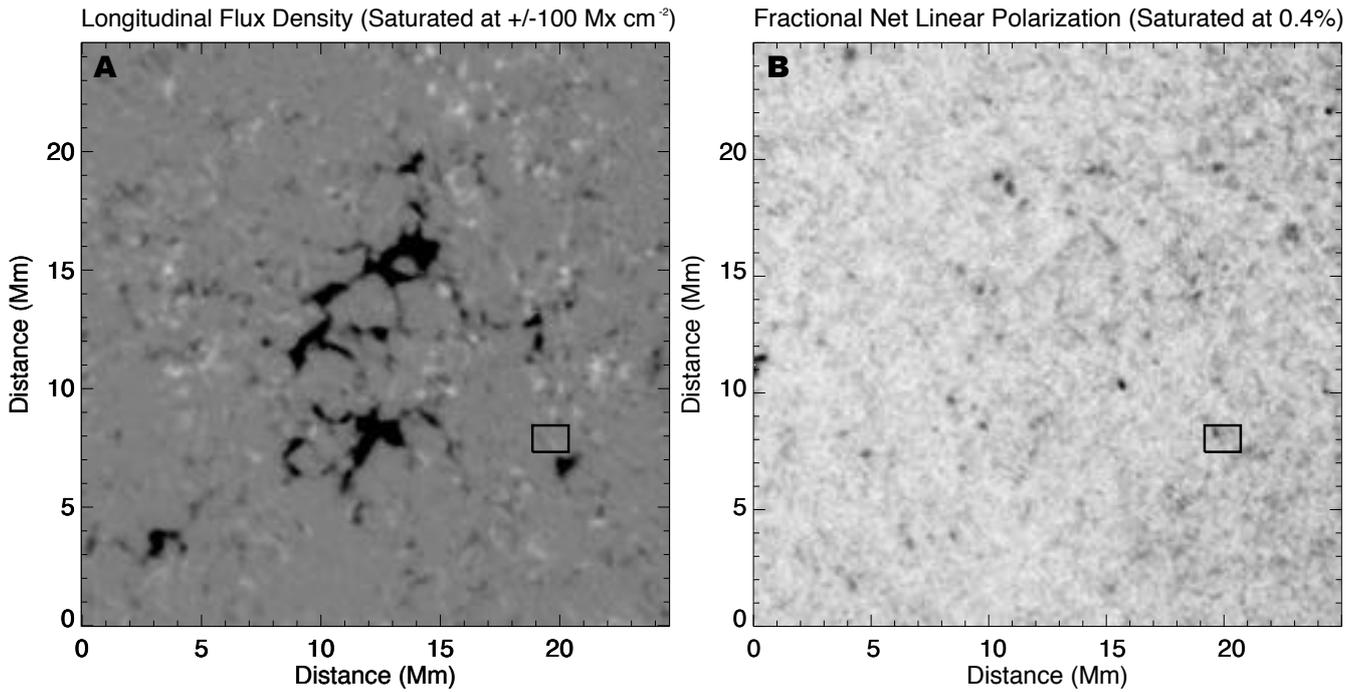

**Fig. S3: Longitudinal flux density and fractional net linear polarization in the ROI at 18:54 UT.** Images are 3-pixel smoothed. In the image of longitudinal flux density (A), negative and positive polarities are shown in white and black, respectively. The linear polarization image (B) is saturated at 0.4%, and the black patches represent stronger signals of linear polarization. The black rectangle in each panel indicates the same region outlined by the black rectangles in Fig. S4.

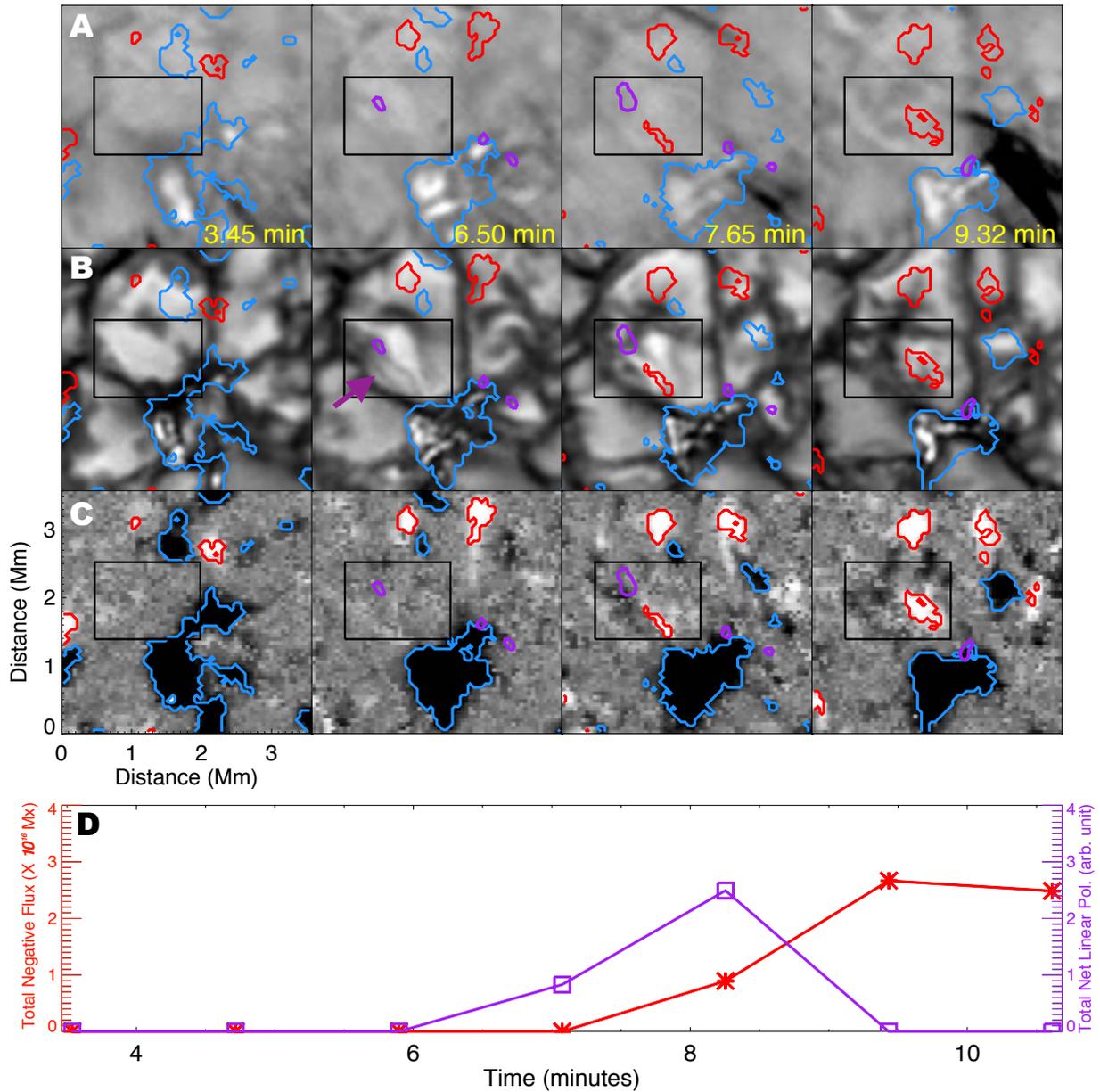

**Fig. S4. Enhanced spicular activity accompanied by the emergence of a bipole/loop.** (A-C) Images of the Hα blue wing, TiO and longitudinal flux density at four different times. Blue and red contour levels are the same as in Fig. 1B. The purple contours show strong linear polarization signals (>0.25%). The purple arrow in (B) indicates the transient darkening of the granule and first appearance of the linear polarization signal. (D) Variations of the total negative flux (stars) inside the red contours and the total net fractional linear polarization (squares) inside the purple contours within the black rectangle shown in (A-C).

T. Samanta, H. Tian, V. Yurchyshyn, et al., Science 366, 890 (2019)14

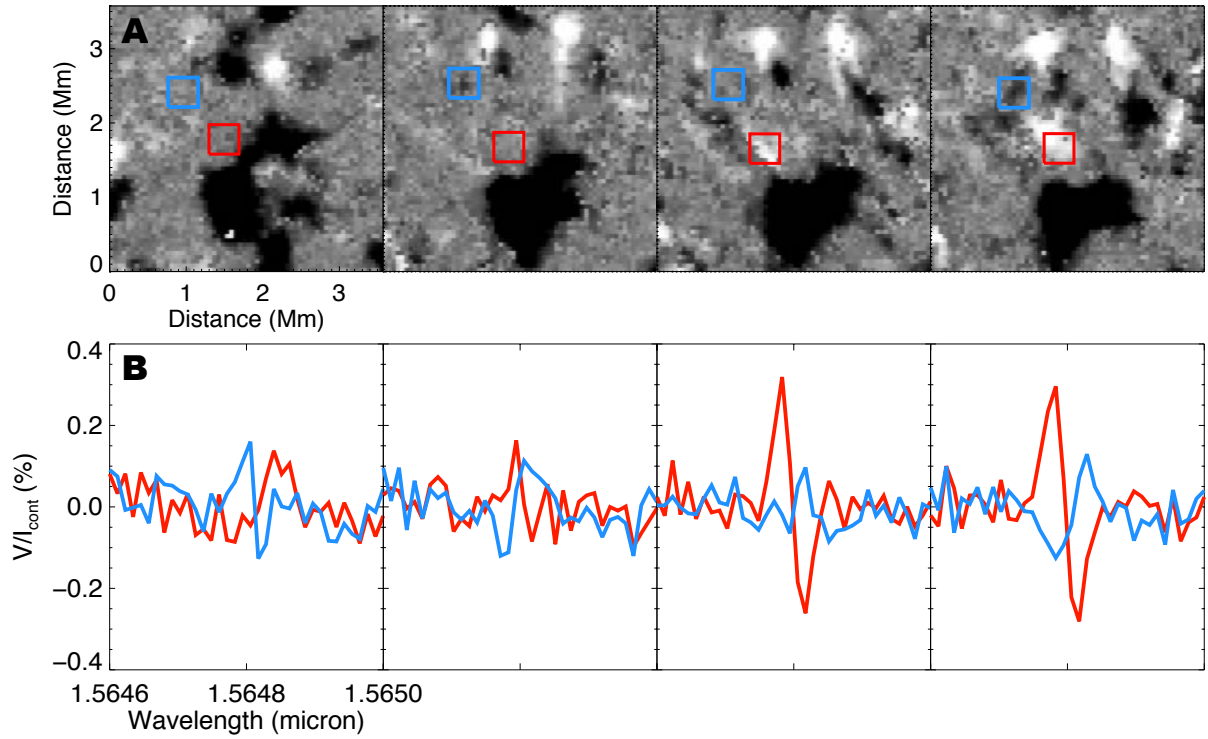

**Fig. S5. Signature of different polarities at the two footpoints of an emerging magnetic bipole/loop (continuation of Fig. S4).** (A) The same as Fig. S4C. (B) The average Stokes-*V* profiles inside the blue and red boxes in (A). The last panel clearly shows opposite polarities for the two footpoints.

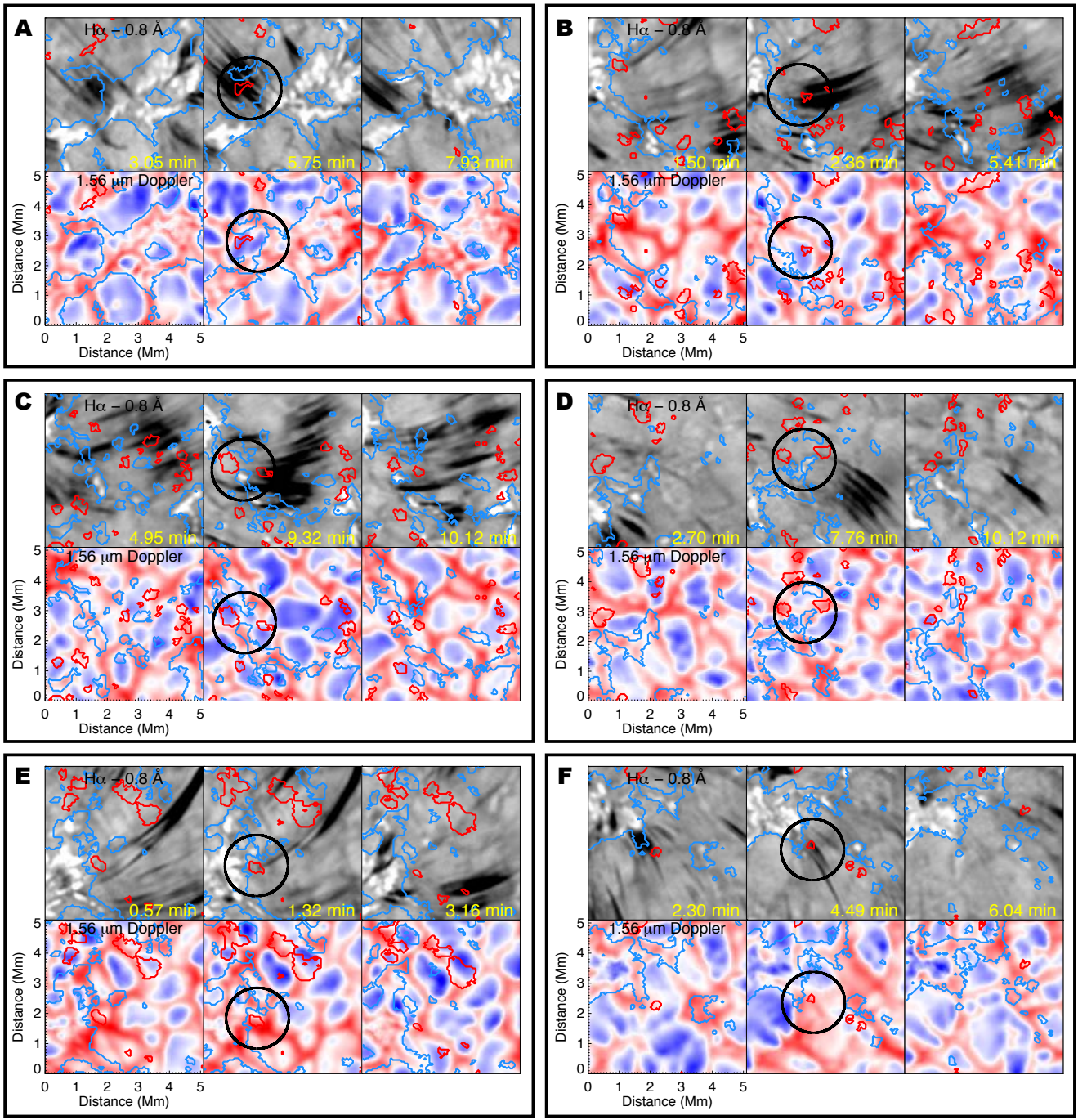

**Fig. S6. Examples of flux emergence/cancellation leading to the generation of spicules.** (A) Temporal evolution of Hα blue wing and Fe I 1.56 μm Doppler shift (saturated at +/- 2 km s$^{-1}$) around the spicule footpoint region (black circle). Contour colors and levels are the same as in Fig. 1B. (B-F) The equivalent data for other examples. Panels (A-D) show examples of the generation of enhanced spicular activities during the stage of flux emergence/appearance. Panels (E-F) show examples of spicule generation during the stage of flux cancellation/decrease.

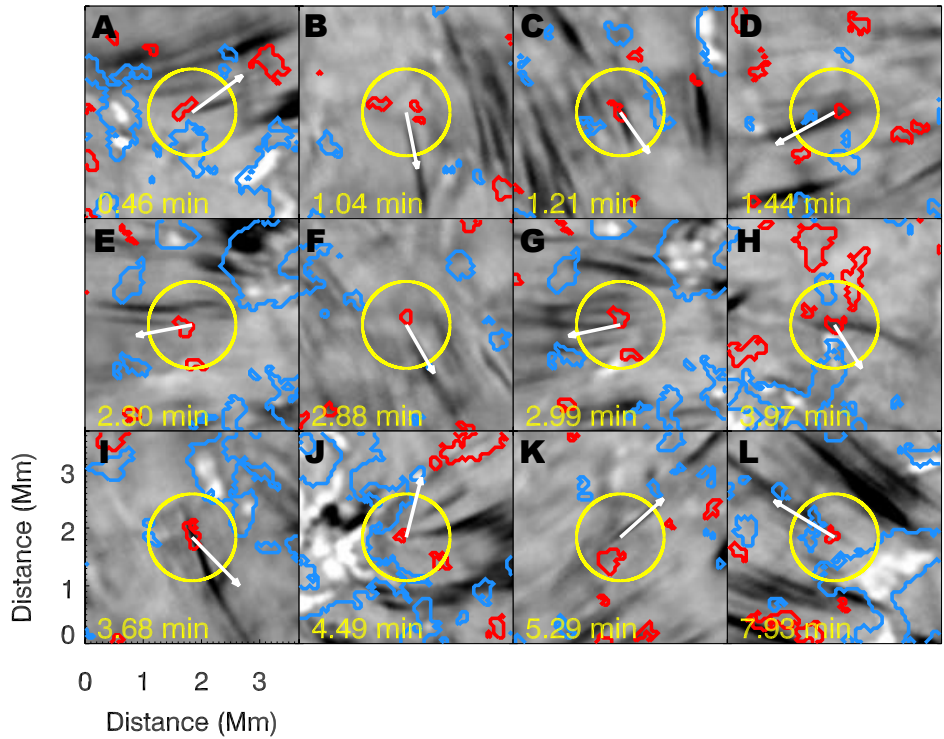

**Fig. S7. Additional examples showing the presence of an opposite-polarity flux near the spicule footpoint.** These examples are presented in the same way as in Fig. 3B.

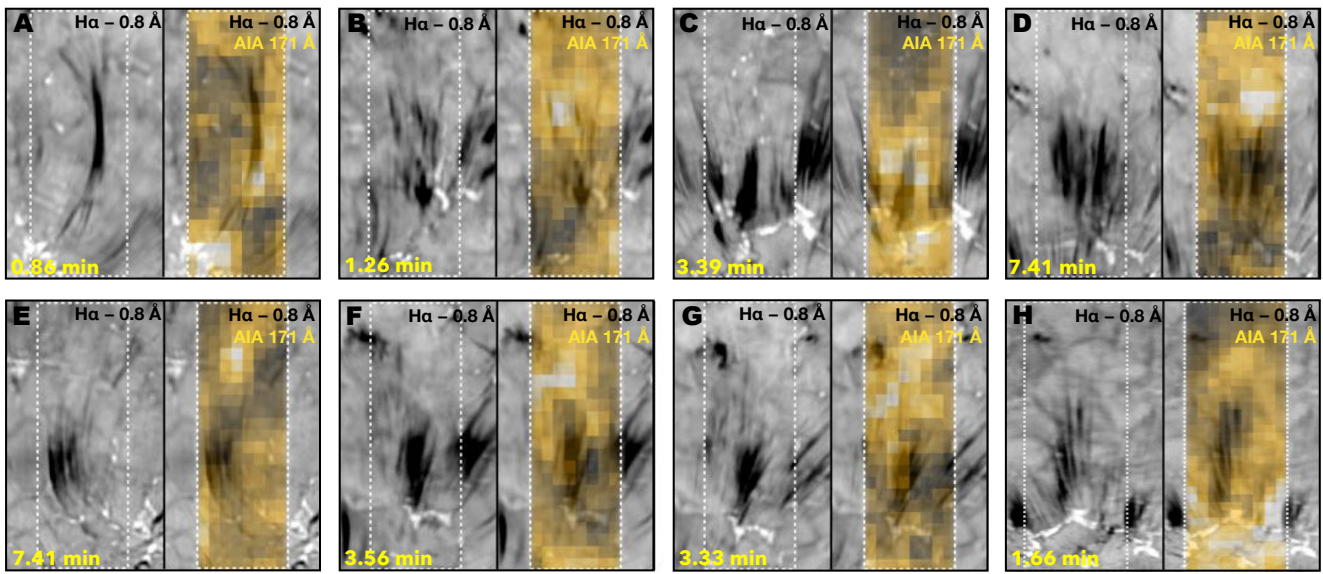

**Fig. S8. Additional eight examples showing the enhancement of coronal emission caused by enhanced spicular activities (extension of Fig. 4).** For each example, the left panel shows an Hα blue wing image. The FOV is 5.5 Mm × 9.5 Mm for (A-G) and 5.5 Mm × 14.5 Mm for (H). The dotted rectangle highlights the enhanced spicular area. The right panel shows the Hα blue wing image overlain with the simultaneously taken AIA 171 Å image.

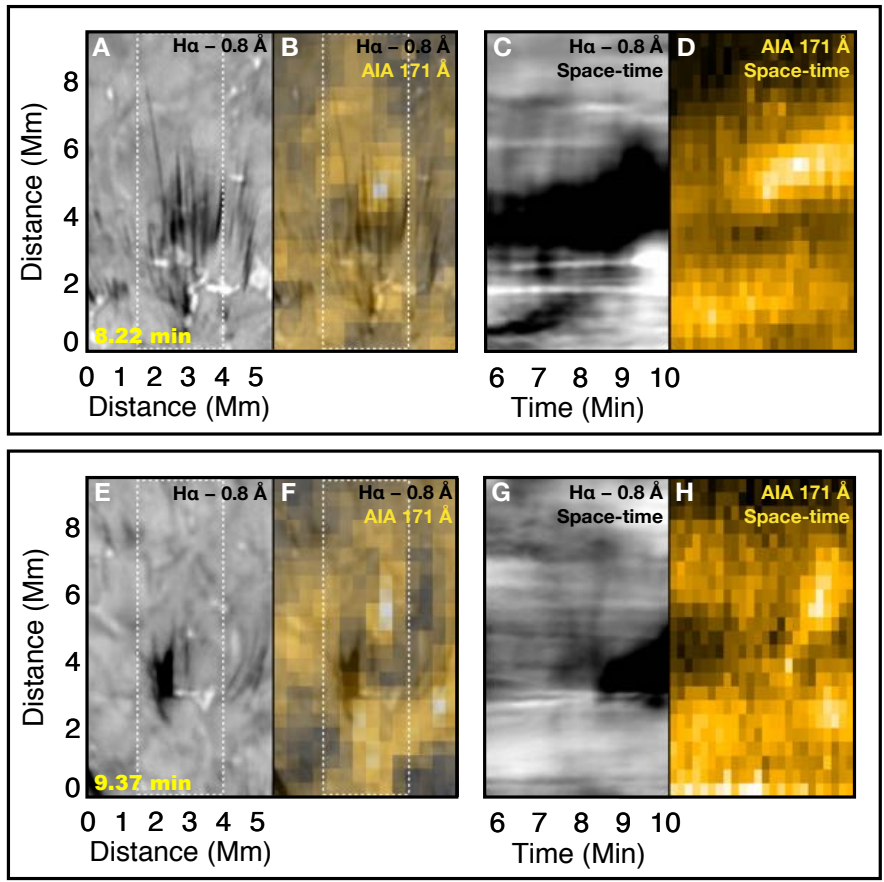

**Fig. S9. Two examples showing heating of spicules.** (A) Hα blue wing, (B) Hα blue wing overlain with the simultaneously taken AIA 171 Å image, (C) space-time plots of Hα blue wing and (D) AIA 171 Å emission for the dashed rectangle shown in panels (A) and (B). Panels (E-H) show the same data for a different event. Movies S7 and S8 show animated versions of (A-D) and (E-H), respectively.

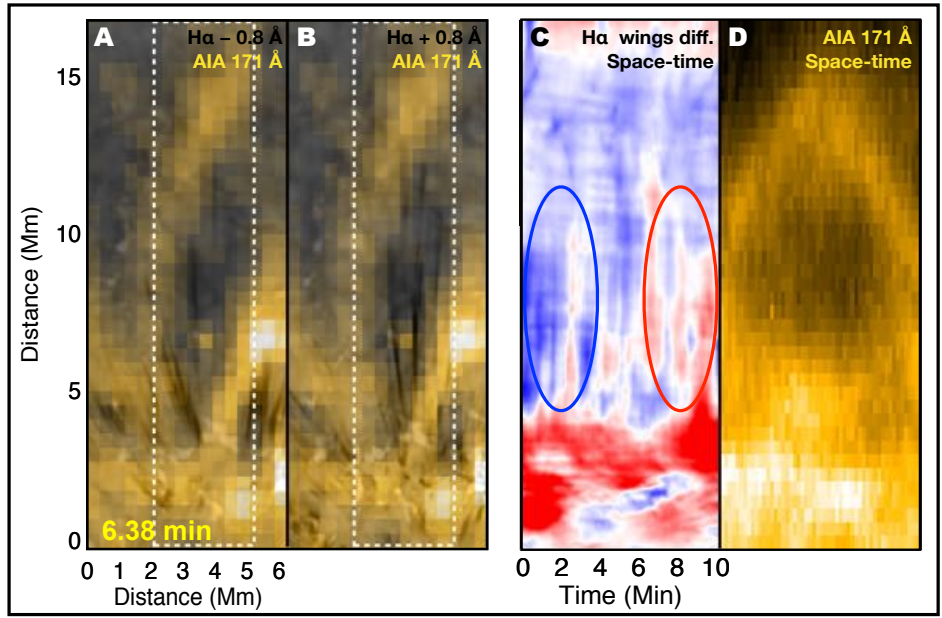

**Fig. S10. Backward motion of spicules.** (A-B) Hα blue and red wing images overlain with the simultaneously taken AIA 171 Å image. (C-D) Space-time plots of the Hα wing difference (blue - red) and AIA 171 Å emission for the dashed rectangle shown in (A) and (B). At the beginning the Hα wing difference indicates a strong blue shift (highlighted by a blue ellipse) due to an upward propagation, later the wing difference is mostly dominated by red shifts (highlighted by a red ellipse) which indicate downflows. The AIA 171 Å space-time diagram shows upward propagation of the coronal counterpart of the spicular feature and subsequent backward motion. Movie S9 shows an animated version of this figure.

## 4. Table

| Date and time | Telescope | Center of ROI (x, y) | Passband or spectral line | Cadence | Pixel size |
|---|---|---|---|---|---|
| 2017 June 19 18:45:58 UT - 18:56:08 UT | GST | (-96", -10") | VIS Hα - 0.8 Å<br>VIS Hα + 0.8 Å | 3.45 s | 21 km |
| | | | NIRIS Fe I 1.56 μm | 71 s | 56 km |
| | | | BFI TiO | 15 s | 25 km |
| | SDO/AIA | (-96", -10") | 171 Å, (1700 Å) | 12 s, (24 s) | 435 km |

**Table S1. Summary of the GST and AIA observations.**
T. Samanta, H. Tian, V. Yurchyshyn, et al., Science 366, 890 (2019)

# 5. Captions for Movies

**Movie S1:** Animated version of Fig. 1A.

**Movie S2:** An example of enhanced spicular activity observed during flux emergence. Panels (A-C): Hα blue wing images, TiO images and photospheric Dopplergrams. Panels (D-F) show a smaller-FOV (white dotted box shown in the top panels) of the same images. Contour colors and levels are the same as in Fig. 1B. The white arrows indicate the location of flux emergence.

**Movie S3**: An example of enhanced spicular activity observed during flux cancellation. Panels (A-C): Hα blue wing images, TiO images and photospheric Dopplergrams. Panels (D-F) show a smaller-FOV (white dotted box shown in the top panels) of the same images. Contour colors and levels are the same as in Fig. 1B.

**Movie S4:** Animated version of Fig. 3A. The footpoint regions of some spicules are marked by the black (enhanced spicular activities including the nine examples shown in Fig. 1B-J and thirteen additional examples) and yellow circles (individual spicules including 16 examples shown in Fig. 3B-Q and 12 examples shown in Fig. S7). Compared to Fig. 3A, a slightly smaller FOV (20.6 Mm × 20.6 Mm) is shown here.

**Movie S5:** Coronal connection of enhanced spicular activities. (A) Hα blue wing images. (B) Hα blue wing images overlain with the simultaneously taken AIA 171 Å images. The rectangular boxes mark 14 examples showing coronal signatures of enhanced spicular activities.

**Movie S6:** Animated version of Fig. S1B-E.

**Movie S7:** Animated version of Fig. S9A-D.

**Movie S8:** Animated version of Fig. S9E-H.

**Movie S9:** Animated version of Fig. S10.